\input mont.sty
\input dc_mont.sty

\hsize=16cm \vsize=21cm
\hfuzz=0.2cm
\tolerance=400
\noindent{\twelverm Gluon Condensate from Superconvergent QCD Sum
 Rule\footnote*{\petit Contribution to Prof. L. Okun's seventieth birthday}}
\vskip0.4cm
\noindent F. J. Yndur\'ain
\vskip0.3cm
\noindent Departamento de F\'\i sica Te\'orica, C-XI,
Universidad Aut\'onoma de Madrid,\hb
Canto Blanco, 28049-Madrid, Spain
\vskip0.6cm

{\petit \noindent Sum rules for the nonperturbative
 piece of correlators (specifically, the vector current 
correlator) are discussed. The sum rule subtracting the perturbative part is of the 
superconvergent type. Thus it is dominated by 
the bound states and the low energy production cross section. It leads to a determination of the 
gluon condensate $\langle\alpha_s G^2\rangle$. We find
$$\langle\alpha_s G^2\rangle\simeq 0.048\pm 0.030\;{\gev}^4.$$}

\begindc{
\vskip-.1cm
\noindent{\fib 1. SUM RULE}
\smallskip
\noindent The potential, or more generally the spectrum of a system of heavy quarks 
cannot be directly discussed in terms of the OPE (operator product expansion).
 However, one can use dispersion relations to deduce a number of sum rules
 relating bound state properties to quantities obtainable {\sl via} the OPE (``ITEP-type"
 sum rules). One 
can then use the estimates of nonperturbative contributions to 
bound states energies and wave functions to actually go beyond the traditional analysis. 
Although the sum rules, being {\sl global} relations, cannot discriminate details 
one can check consistency and even obtain reasonable estimates on nonperturbative 
quantities, specifically on the gluon condensate. This last is 
the aim of the present note, where we will use a method generalizing that 
proposed by Novikov\ref{1}.

To do so we consider the correlator for the vector current of heavy quarks,
$$\eqalign{\Pi_{\mu\nu}=&(p^2g_{\mu\nu}-p_{\mu}p_{\nu})\Pi(p^2)\cr
=&
{\ii}\int \dd ^4 x\,\ee ^{\ii p\cdot x}\langle {\rm T}J_{\mu}(x)J_{\nu}(0)\rangle\,,\cr}
\eqno (1)$$
where $J_{\mu}=\bar{\psi}\gamma_{\mu}\psi$ and sum over omitted colour indices 
is understood. This will give information on triplet, $l=0$ states; information 
on states with other quantum numbers would be obtained with other correlators. 
The function $\Pi (t)$ satisfies a dispersion relation,
$$
\Pi (t)={{1}\over{\pi}}\int \dd s{{\rho (s)}\over{s-t}}\,,
\eqno (2)$$
where $\rho (s)\equiv \imag \Pi (s)$.
Actually, this equation should have been written with one subtraction. We will 
not bother to do so as its contribution drops out for
 the quantities of interest for us here.

Let us denote by $\Pi_{\rm p.t.},\,\rho_{\rm p.t.}$ to the corresponding 
quantities calculated in perturbation theory, albeit {\sl to all orders}, 
but nonperturbative effects are neglected 
in  $\Pi_{\rm p.t.},\,\rho_{\rm p.t.}$.  (In 
actual calculations we cannot of course include {\sl all} orders. 
We will sum the one-gluon exchange to all orders (which 
can be done explicitly in the nonrelativistic regime), and  add one loop  
radiative corrections to this.)  In particular, for example, 
the gluon condensate contribution is not included in the ``p.t." pieces. 

At large $t$, both spacelike and timelike, the OPE is applicable to $\Pi (t)$, 
and we have the well-known results\ref{2},
$$
\Pi (t)\simeq\Pi_{\rm p.t.}(t)+{{\langle\alpha_s G^2\rangle}\over{12\pi t^2}}
\eqno (3)$$
and
$$
\rho (s)\simeq\rho_{\rm p.t.}(s)-{{N_cC_F}\over{128}}\,
{{\langle\alpha_s G^2\rangle}\over{s^2}}\,{{(1+v^2)(1-v^2)^2}\over{v^5}}
\eqno (4)$$
with $v=(1-4m^2/s)^{{1}\over{2}}$ the velocity of the quarks. 
Moreover,
$$\eqalign{\Pi_{\rm p.t.}(t)\simeqsub_{t\to\infty}&\,-\dfrac{N_c}{12\pi^2}
\Bigg\{\log\dfrac{-t}{\nu^2}\cr
+&\dfrac{3C_F}{\beta_0}\log\log\dfrac{-t}{\nu^2}+\cdots\Bigg\},
\cr
\imag\Pi_{\rm p.t.}(s)\simeqsub_{t\to\infty}&\,
\dfrac{N_c}{12\pi}\left\{1+\dfrac{3C_F\alpha_s}{4\pi}+\cdots\right\};\cr
 N_c=&\,3,\;C_F=4/3.\cr}$$
 If we then define 
$\Pi_{\rm NP}\,,\rho_{\rm NP}$ as the results of subtracting the perturbative parts,
 $$
\Pi_{\rm NP}\equiv \Pi-\Pi_{\rm p.t.}\,;\;\rho_{\rm NP}\equiv\rho-\rho_{\rm p.t.},
$$
it follows from the OPE, \equ~(3), that $\Pi_{\rm NP}$ decreases at infinity 
like $t^{-2}$ and hence satisfies a superconvergent dispersion relation. We thus
 have a first sum rule:
$$
\int \dd s\,\rho_{\rm NP}(s)=0.\eqno (5)$$
In fact it would appear that one still has another sum rule because of the following 
argument. At large $t$, $\Pi_{\rm NP}(t)$ behaves like (cf. Eq. (3))
$$\Pi_{\rm NP}(t)\simeq {{\langle\alpha_s G^2\rangle}\over{12\pi t^2}}, $$
while the contribution from the bound states to the dispersion relation (see below), 
$$\Pi_{\rm NP; bound\;states}(t)\sim {{\langle\alpha_s G^2\rangle}\over{ t^2\alpha_s^3}}
\eqno (6)$$ 
dominates over this. Therefore we have the extra relation,
$$\int \dd s\,s\rho_{\rm NP}(s)=0.\eqno (7)$$
It turns out that (7) is actually equivalent to (5), up to radiative corrections. This is 
because the region where any of the integrals in (5), (7) are appreciably different 
from zero is for $s\simeq 4m^2 (1+{\rm O}(\alpha_s^2))$, so (7) differs from (5) 
by terms of order $\alpha_s^2$, smaller than the radiative corrections which 
neither (5) nor (7) take into account.

Let us return to the sum rule (5). The function 
$\rho(s)$ consists of a 
continuum part, for $s$ above threshold for open bottom production, and a sum of 
bound states. Both can be calculated theoretically 
provided that $s$ is larger than a certain critical $s(v_0)$, and $n$ smaller or 
equal than a critical $n_0$. $s(v_0)$ and $n_0$ are defined 
as the points where the perturbation theoretic contribution to $\rho$ and
 the nonperturbative one are of equal magnitude, and form the limits 
of the regions where a full theoretical evaluation is possible. 

To be precise, 
for the continuum we use (4) so that above the critical $s(v_0)$,
$$\eqalign{\rho^{\rm cont}_{\rm NP}(s)=&{{N_cC_F}\over{128}}\,
{{\langle\alpha_s G^2\rangle}\over{s^2}}\,{{(1+v^2)(1-v^2)^2}\over{v^5}},\cr
s>s(v_0),\cr}
\equn{(8a)}$$
and $v_0$ is such that
$\rho^{\rm cont}_{\rm NP}(s(v_0))\simeq\rho^{\rm cont}_{\rm p.t.}(s(v_0))$;
numerically, and for $\bar{b}b$, $v_0\simeq 0.2$.
For the bound states $\rho$ is proportional to the square of the 
wave function at the origin,
$$\rho(s)=\dfrac{N_c}{M_n}\,|R_{n}(0)|^2\delta(s-M_n).$$
We may get $\rho^{\rm b.s}_{\rm p.t.}(s)$ and $\rho^{\rm b.s.}_{\rm NP}(s)$ 
by splitting the residue $|R_{n}(0)|^2$ into a Coulombic piece,
$$|R_{n}^{\rm Coul.}(0)|^2=\dfrac{m^3C_F^3\alpha_s^3}{2n^3}(1-\delta^{\rm b.s.}_n\alpha_s),$$
where the one loop corrections $\delta^{\rm b.s.}_n\alpha_s$ may be found in ref. 3, 
and the (leading) nonperturbative correction are given by the Leutwyler--Voloshin 
analysis (refs. 4, 3). So we have
$$|R_{n}(0)|^2\simeq|R_{n}^{\rm Coul.}(0)|^2+|R_{n}^{\rm Coul.}(0)|^2\delta^{\rm NP}_n; $$
the numbers $\delta^{\rm NP}_n$ have been calculated by Leutwyler and Voloshin. 
For $n=1$, 
$$\delta^{\rm NP}_1=\dfrac{38.3\langle\alpha_sG^2\rangle}{m^3C_F^2\alpha_s^2}.$$
 This is all we 
really need since, for bottomium, $n_0=1$. Thus we have 
$$\eqalign{\rho^{\rm b.s.}_{\rm NP}(s)=&{{3N_c C_F^3 \pi m^3\,\langle \alpha_s G^2\rangle}
\over{8\alpha_s^3 m^4}}\sum^{n_0}_{n=1}{{\eta_n }\over{M_n}}\delta(s-M^2_n),\cr
n\leq& n_0,\cr}
\equn{(8b)}$$
the $\eta_n$ known in terms of the $\delta^{\rm NP}_n$.

The sum rule (5) can then be written schematically as
$$\eqalign{\int_{s(v_0)}^\infty \rho_{\rm NP}+\sum_{n=1}^{n_0}{\rm Residue\;of}\,\rho_{NP}\cr
=\left\{\int^{s(v_0)}_{\rm threshold} \rho_{\rm NP}
+
\sum_{n=n_0+1}{\rm Residue\;of}\,\rho_{NP}\right\}.\cr}$$
The left hand side is given in terms of $\langle\alpha_s G^2\rangle$ 
by \equs~(8); the right hand side can
be connected with experiment with the 
following argument. The sum 
over higher bound states,
 
``$\sum_{n=n_0+1}{\rm Residue\;of}\,\rho_{NP}$"

\noindent may be identified as the difference between the
sum over the {\sl experimental} residues of the poles of the bound states, and what we would get 
by a Coulombic formula, for all $n\geq n_0+1$. Certainly, this Coulombic formula will not be 
valid for large $n$ because here the radiative corrections will become 
large; but, because the residues decrease like $1/n^3$ the contribution of these states 
will be negligible. We write this decomposition as
$$\eqalign{({\rm bound\;states\;with}\;\,n>n_0)=&\rho^{\rm b.s.}_{{\rm exp},\,n>n_0}(s)\cr
-&\rho^{\rm b.s.}_{{\rm Coulombic},\,n>n_0}(s).\cr}$$ 
As for the continuum piece below $s(v_0)$ we may likewise interpret it as the difference between 
experiment and a perturbative evaluation, which we write as
$$\rho^{\rm cont}_{\rm NP}(s)=
\rho^{\rm cont}_{\rm exp}(s)-\rho^{\rm cont}_{\rm p.t.}(s),\quad s<s(v_0),$$
and, because we are close to threshold, we have 
$$\eqalign{\rho^{\rm cont}_{\rm p.t.}(s)=&{{N_cC_F\alpha_s}\over{8}}\;
{{1}\over{1-\ee^{-\pi C_F\alpha_s/v}}}\left(1+\delta_{\rm cont}\alpha_s\right)\cr
\simeq& {{N_cC_F\alpha_s}\over{8}}\left(1+\delta_{\rm cont}\alpha_s\right)\cr}$$
and the value of the one loop radiative correction $\delta_{\rm cont}\alpha_s$
 may be found in ref.~5.

Taking everything into account the sum rule (5) becomes,
$$\eqalign{\sum_{n=n_0+1}{{1}\over{m^2M_n}}|R^{\rm exp}_n(0)|^2+f_{\rm back}(v_0)=\cr 
2C_F^3\left\{\alpha_s^3\sum^{\infty}_{n=n_0+1}{{1}\over{n^3}}-
{{\pi\langle \alpha_s G^2\rangle}\over{m^4\alpha_s^3}}
\sum_{n=1}^{n_0}\lambda_{n} n^5\right\}\cr
+{\textstyle{{2}\over{3}}}\left\{8\epsilon^2\alpha_s^3+
{{\langle \alpha_s G^2\rangle}\over{48\epsilon^3\alpha_s^3 m^4}}\right\}.}\equn{(9)}$$
We have defined $v_0\equiv\epsilon\alpha_s$ and the expression is valid 
up to corrections of relative order $\alpha_s$. The function $f_{\rm back}(v_0)$ is the 
contribution of the background which, when added to the resonances above threshold (included 
in the sum in the l.h.s. of (9)), give the experimental value of
 $\int^{s(v_0)}_{\rm threshold} \rho_{\rm NP}$.
The function $f_{\rm back}$ would be obtained by integrating the cross sections 
for production of $\Upsilon+G$ and $B\bar{B}$, where by $G$ we mean a ``glueball" decaying 
into $2\pi$, and $B$ is any of the states $B^0,B^{\pm}, B^*$. Because we may assume that 
the structure is provided by the resonances, we can take $f_{\rm back}$ given by phase space 
only. So we have
$$f_{\rm back}(v_0)=f_1 v_0^{5/2}+f_2 v_0^3 $$
where the first term refers to the channel $\Upsilon+G$ , and the second to $B\bar{B}$. 
We have in this expression neglected $m_G$.

\noindent{\fib 2. NUMEROLOGY}

\noindent In principle the procedure would appear straightforward. One would fit the 
resonance and bound state residues and $f$ to 
the data, and then, after substituting into (9), obtain  a 
determination of $\langle\alpha_s G^2\rangle$. In practice, however, things 
do not work out so nicely. The quality of the experimental data does not 
allow any precise determination of the constants $f_{1,2}$; any values 
in the range
$f_{1,2}\sim 0.03,\,0.1$
would do the job. Secondly, the effective dependence of $\langle\alpha_s G^2\rangle$ 
in \equ~(9) on experiment is proportional to $\alpha_s^{-6}$: so
 the result will depend very strongly on 
the value of $\alpha_s$  we choose. This is particularly true because radiative 
corrections to the nonperturbative contribution to the 
bound states have not been calculated, so there is not even a
 ``natural" renormalization point.

These two difficulties may be partially overcome with the following tricks. First of all, 
since we are assuming that the $n=1$ bound state is described with the bound 
state analysis as discussed in ref.~3,  
 we may fix the value of $\alpha_s$ that produces such agreement.
 This means that we will take $0.35\leq \alpha_s\leq 0.4.$ Secondly, we may alter
 the treatment of the continuum in the following manner. We split {\sl not} from $v_0$, but from 
$v_1$, arbitrary provided only that $v_1\geq v_0$. Thus, for $s\leq s(v_1)$, we use
$\rho^{\rm cont}_{\rm NP}(s)=
\rho^{\rm cont}_{\rm exp}(s)-\rho^{\rm cont}_{\rm p.t.}(s)$, and for $s\geq s(v_1)$ we take the 
theoretical expression
$$\rho^{\rm cont}_{\rm NP}(s)={{N_cC_F}\over{128}}\,
{{\langle\alpha_s G^2\rangle}\over{s^2}}\,{{(1+v^2)(1-v^2)^2}\over{v^5}}.$$
The sum rule is thus written as
$$\eqalign{\sum_{n=2}{{1}\over{m^2M_n}}&
|R^{\rm exp}_n(0)|^2+f_{\rm back}(v_1)\cr
=&2C_F^3\left\{[\zeta(3)-1]\alpha_s^3
-{{4.9\,\langle\alpha_s G^2\rangle}\over{m^4\alpha_s^3}}\right\}\cr
+&{\textstyle{{2}\over{3}}}\left\{8\epsilon_1^2\alpha_s^3+
{{\langle \alpha_s G^2\rangle}\over{48\epsilon_1^3\alpha_s^3 m^4}}\right\},\quad
\epsilon_1\alpha_s=v_1.\cr}$$   
Then we may profit from the fact that the sum rule should be valid for {\sl all}  values
 of $v_1\geq v_0$ to fix $f_{1,2}$ requiring this independence, at least in the mean. That is 
to say, that when we increase $v_1$ past a particle threshold from $\Upsilon (2)$
 to $\Upsilon (6)$ the variation of the corresponding determinations of
 $\langle\alpha_s G^2\rangle$ around their average be minimum. The calculation 
may be further simplified replacing 
$$f_{\rm back}(v_1)\to2f_0v_1^{2.75}.$$
The results of the analysis are summarized in the following tables, where 
the column ``Res" indicates at which resonance the cut in $v_1$ occurs.
We have taken two rather extreme values of $f_0$.
\vskip.2cm
\setbox2=\vbox{\hsize 5.5cm
\setbox1=\vbox{\offinterlineskip\hrule
\halign{
&\vrule#&\strut\hfil#\hfil&\quad\vrule\quad#&\strut\quad#\quad&\quad\vrule#&\strut\quad#\cr
 height2mm&\omit&&\omit&&\omit&\cr 
& \kern.5em{\rm Res.}&&$v_1$&& $\langle\alpha_s G^2\rangle$\kern.3em& \cr
 height1mm&\omit&&\omit&&\omit&\cr
\noalign{\hrule} 
height1mm&\omit&&\omit&&\omit&\cr
&  $\Upsilon(2)$&&0.21&&0.014& \cr
&  $\Upsilon(3)$&&0.34&&0.034& \cr
&  $\Upsilon(4)$&&0.40&&0.048&\cr
&  $\Upsilon(5)$&&0.43&&0.039&\cr
&  $\Upsilon(6)$&&0.46&&0.046& \cr
 height1mm&\omit&&\omit&&\omit&\cr
\noalign{\hrule}}
\vskip.05cm}
\centerline{\box1}
{\petit
\centerline{For $\alpha_s=0.35, f_0=0.04$}}
\centerrule{1cm}
\vskip.3cm}
\setbox3=\vbox{\hsize5.5cm
\setbox0=\vbox{\offinterlineskip\hrule
\halign{
&\vrule#&\strut\hfil#\hfil&\quad\vrule\quad#&\strut\quad#\quad&\quad\vrule#&\strut\quad#\cr
 height2mm&\omit&&\omit&&\omit&\cr
& \kern.5em{\rm Res.}& &$v_1$& & $\langle\alpha_s G^2\rangle$\kern.3em&\cr
 height1mm&\omit&&\omit&&\omit&\cr
\noalign{\hrule}
 height1mm&\omit&&\omit&&\omit&\cr
& $\Upsilon(2)$& &0.21& &0.037&\cr
& $\Upsilon(3)$& &0.34& &0.057&\cr
& $\Upsilon(4)$& &0.40& &0.067&\cr
& $\Upsilon(5)$& &0.43& &0.048&\cr
& $\Upsilon(6)$& &0.46& &0.052&\cr
 height1mm&\omit&&\omit&&\omit&\cr
\noalign{\hrule}}
\vskip.05cm}
\centerline{\box0}
{\petit
\centerline{For $\alpha_s=0.40, f_0=0.09$}}
\centerrule{1cm}
\vskip.3cm}
\centerline{\box2}
\medskip
\centerline{\box3}

This derivation shows very clearly the kind of 
errors one encounters. To the variations that may be called ``statistical",
 apparent in the different 
values found in the tables above
$$0.014\leq\langle\alpha_s G^2\rangle\leq0.067$$
we have to add ``systematic" ones, e.g., the influence of 
the not calculated radiative corrections, easily of 
some 30\%: not to mention our including the Coulombic wave functions at the origin 
for large values of $n$, or the lack of definition of the 
expression ``perturbation theory to all orders" because of renormalon ambiguities. 
 Given all these uncertainties, which do even make it dubious that one can 
really define with precision the condensate in terms of experimental observables,
 it is not surprising that one 
cannot pin down the gluon condensate with more accuracy than 
an estimate, taking into account above figures, of 
$$\langle\alpha_s G^2\rangle\simeq 0.048\pm 0.03\,{\gev}^4.$$
To get this average we have taken into account all determinations in the 
 tables above, excluding the lowest ($\Upsilonv(2)$) and highest, $\Upsilonv(6)$.
This is slightly larger than old averages,
 and slightly lower than more recent ones\ref{6} which tended to give, respectively,
$\langle\alpha_s G^2\rangle\simeq 0.042,\;\langle\alpha_s G^2\rangle\simeq0.065\,{\gev}^4.$

\brochuresubsection{ACKNOWLEDGEMENTS}

\noindent Discussions with R. Akhoury and V. Zakharov 
on some aspects of the sum rule are gratefully acknowledged.
      
\brochuresubsection{REFERENCES}
{\petit
\item 1.- See, e.g., V. A. Novikov et al., Phys. Rep. {\bf C41} (1978), 1.
\item 2.- M. A. Shifman, A. I. Vainshtein and V. I. Zakharov,
Nucl. Phys. {\bf B147} (1978) 385.
\item 3.- S. Titard and F. J. Yndur\'ain, Phys. Rev. {\bf D49} (1994), 6007 and 
{\bf D51} (1995), 6348; A. Pineda and F.~J.~Yndur\'ain,
 Phys. Rev., {\bf D58} (1998), 094022 and CERN-TH/98-402 (hep-ph/9812371).
\item 4.- H. Leutwyler, Phys. Lett. {\bf B98} (1981) 447; 
M. B. Voloshin Nucl. Phys. {\bf B154} (1979) 155; Sov. J. Nucl. Phys.
{\bf 36} (1982) 143.
\item 5.- K. Adel and F. J. Yndur\'ain, Phys. Rev. {\bf D52} (1995), 6577.
\item 6.- Cf. the reviews of S. Narison, {\sl QCD Spectral Sum Rules}, World Scientific, 1989 
and Nucl Phys. Suppl. {\bf 54A} (1997), 238.
\item{}{}
}
}\enddc

\bye